\documentclass[final,5p,times,twocolumn]{elsarticle}

\usepackage{graphicx}
\usepackage{dcolumn}
\usepackage{bm}

\usepackage{mathptmx, courier, pifont}
\usepackage[scaled=0.92]{helvet}
\usepackage[T1]{fontenc}
\usepackage{textcomp}
\usepackage{color}
\usepackage{colordvi}

\newcommand{\quarterthin}{\kern 0.0417em}

\begin{document}
\begin{frontmatter}


\title{{\bf Isospin-symmetry breaking in superallowed Fermi $\beta$-decay
due to isospin-nonconserving forces}}

\author{K.~Kaneko}\address{
Department of Physics, Kyushu Sangyo University, Fukuoka 813-8503,
Japan}

\author{Y.~Sun}\address{School of
Physics and Astronomy, Shanghai Jiao Tong University, Shanghai
200240, China} \address{Collaborative Innovation Center of IFSA,
Shanghai Jiao Tong University, Shanghai 200240, China}
\address{Institute of Modern Physics, Chinese Academy of Sciences,
Lanzhou 730000, China}

\author{T.~Mizusaki}\address{Institute of Natural Sciences, Senshu
University, Tokyo 101-8425, Japan}

\author{S.~Tazaki}\address{Department of Applied Physics, Fukuoka
University, Fukuoka 814-0180, Japan}

\author{S.~K.~Ghorui}\address{School of Physics and Astronomy, Shanghai Jiao Tong
University, Shanghai 200240, China} \address{Collaborative
Innovation Center of IFSA, Shanghai Jiao Tong University, Shanghai
200240, China}


\begin{abstract}
We investigate isospin-symmetry breaking effects in the $sd$-shell
region with large-scale shell-model calculations, aiming to
understand the recent anomalies observed in superallowed Fermi
$\beta$-decay. We begin with calculations of Coulomb displacement
energies (CDE's) and triplet displacement energies (TDE's) by adding
the $T=1,J=0$ isospin nonconserving (INC) interaction into the usual
isospin-invariant Hamiltonian. It is found that CDE's and TDE's can
be systematically described with high accuracy. A total number of
122 one- and two-proton separation energies are predicted
accordingly, and locations of the proton drip-line and candidates
for proton-emitters are thereby suggested. However, attempt to
explain the anomalies in the superallowed Fermi $\beta$-decay fails
because these well-fitted $T=1,J=0$ INC interactions are found no
effects on the nuclear matrix elements. It is demonstrated that the
observed large isospin-breaking correction in the $^{32}$Cl
$\beta$-decay, the large isospin-mixing in the $^{31}$Cl
$\beta$-decay, and the small isospin-mixing in the $^{23}$Al
$\beta$-decay can be consistently understood by introducing
additional $T=1,J=2$ INC interactions related to the $s_{1/2}$
orbit.
\end{abstract}

\end{frontmatter}


The degeneracies of energy levels in nuclei with interchanging
number of protons and neutrons indicate the existence of isospin
symmetry \cite{Heisenberg32,Wigner37}. The concept of this
approximate symmetry is successful in describing various
observables, there are conditions where it does not hold. Isospin
symmetry is broken in QCD due to the mass difference between the up
and down quarks and their electromagnetic interaction
\cite{Miller06}. In nuclei, the Coulomb interaction and the
charge-dependent nucleon-nucleon interaction break this symmetry,
giving rise to observable effects. Especially for nuclei near the
$N=Z$ line, both the ground states and excited spectra are affected by
isospin-symmetry breaking (ISB). Therefore, investigation of
proton-rich nuclei far from the line of stability provides important
testing ground for ISB effects. It has been known that ISB in
nuclear many-body systems in terms of the isospin nonconserving
(INC) interactions \cite{Ormand89} leads to non-zero Coulomb
displacement energy (CDE) \cite{Jaeneke66,Bentley07}
and triplet displacement energy (TDE) (see definitions in Eq.
(\ref{eq:1}) below).
While the microscopic origin of possible INC sources is yet
to be understood, study of these quantities must involve the
knowledge of the many-body effects in nuclear structure
\cite{Bentley12}. This suggests that the detailed shell-model
calculations with an inclusion of INC forces are essential for
further refining our knowledge about ISB. Recently, it has been
shown that the INC interaction of the $T=1,J=0$ channel plays an
important role in the explanation of characteristic behavior of the
$f_{7/2}$-shell CDE's \cite{Brown02,Kaneko13} and TDE's
\cite{Kaneko13}. Mirror energy differences (MED's) \cite{Bentley07}
and triplet energy differences (TED's) \cite{Zuker02,Garrett01} were
previously studied in the $f_{7/2}$-shell, and have recently been
investigated by us for the upper $fp$-shell nuclei \cite{Kaneko12}.
On the other hand, for light nuclei the MED's have
been discussed in the relation with the loosely bound $s_{1/2}$
proton \cite{Thomas52,Ehrman51}, which reduces the Coulomb repulsion
and hence strongly influences the MED's of the corresponding nuclei
\cite{Ogawa99,Yuan14}. We remark that the MED's in the $sd$-shell could have a different
origin from that in the $fp$-shell.

Superallowed Fermi $\beta$-decay serves as a crucial nuclear input
to test the precise values of the Cabbibo-Kobayashi-Maskawa (CKM)
mixing matrix element $V_{ud}$ between the $u$ and $d$ quarks
\cite{Hardy05,Towner08}, provided that the radiative and ISB effects
are considered properly \cite{Marciano06}. In most experiments and
theories, the correction due to ISB is known to be smaller than $2
\%$ \cite{Towner08,Hyland06,Satula11}, and the nucleus-independent
${\mathcal F}t$ values are found to be consistent \cite{Towner08}.
However, recent data have indicated an anomalously large ISB effect
in the superallowed Fermi $\beta$-decay of $^{32}$Cl
\cite{Melconian11} and $^{31}$Cl \cite{Bennett16}. An unusually
large correction $\delta_{C}=5.3(9)\%$ has been reported for the
Fermi $\beta$-decay from $^{32}$Cl to the $T=1$ isobaric analog
state (IAS) in $^{32}$S, and a large isospin mixing has been
observed for the $\beta$-decay from $^{31}$Cl to the $T=3/2$ IAS in
$^{31}$S. It should be noted that these results may depend on both
the INC force and the location of the states that mix. An accidental
near-degeneracy of the states could enhance isospin mixing, without
requiring a particularly strong INC force. This can be estimated by using
the perturbation theory, which, subsequently, can be applied to extract
the INC mixing matrix element from the experimental data of the superallowed
Fermi $\beta$-decays \cite{Tripathi13}. Furthermore, the observation of
the strong isospin mixing for $A=23$ is still controversial
\cite{Iacob06,Tighe95}. A consistent understanding of all the above
requires detailed shell-model calculations for the $sd$-shell
region.

\begin{figure*}[t]
 \begin{center}
\includegraphics[totalheight=8.5cm]{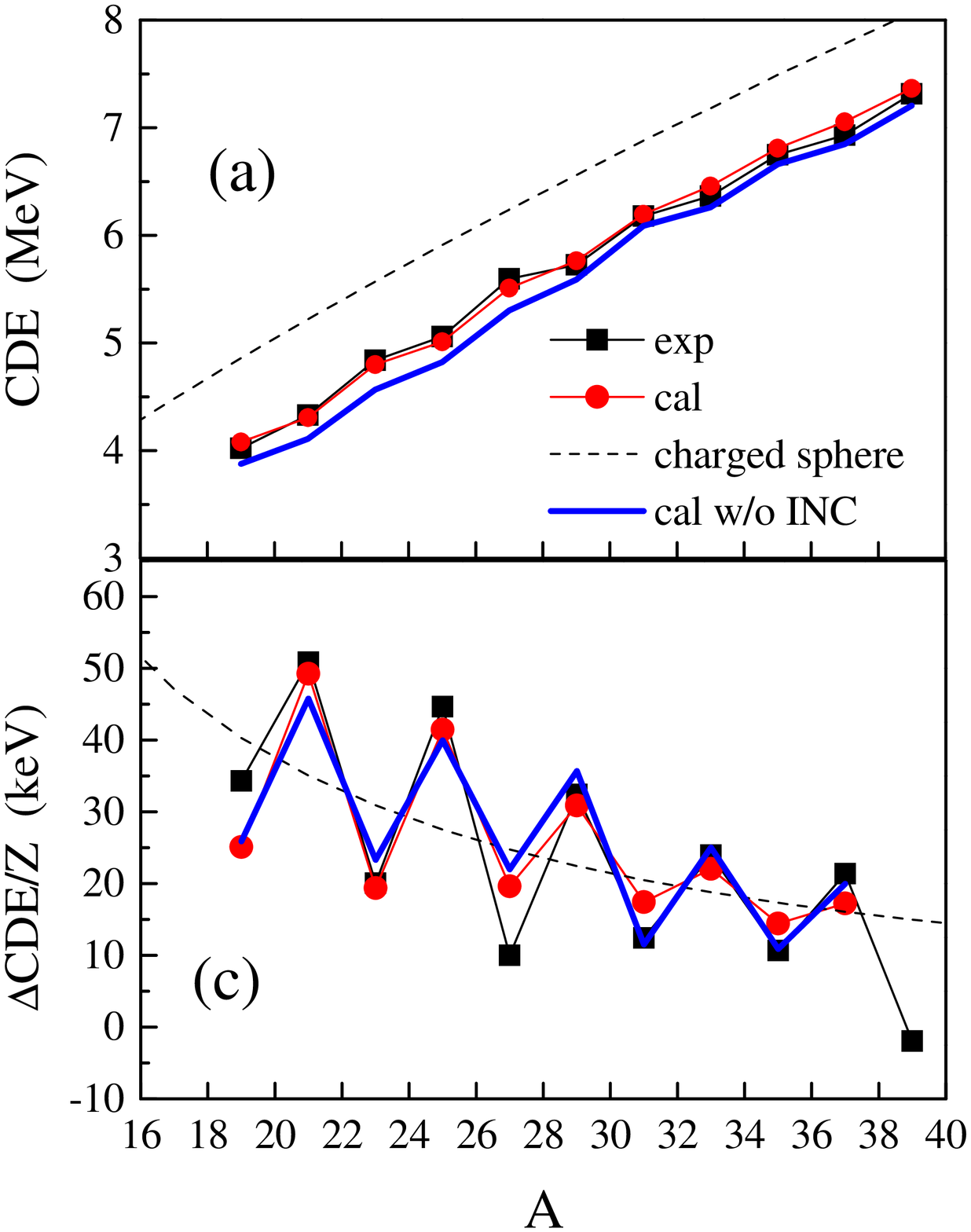}
\includegraphics[totalheight=8.5cm]{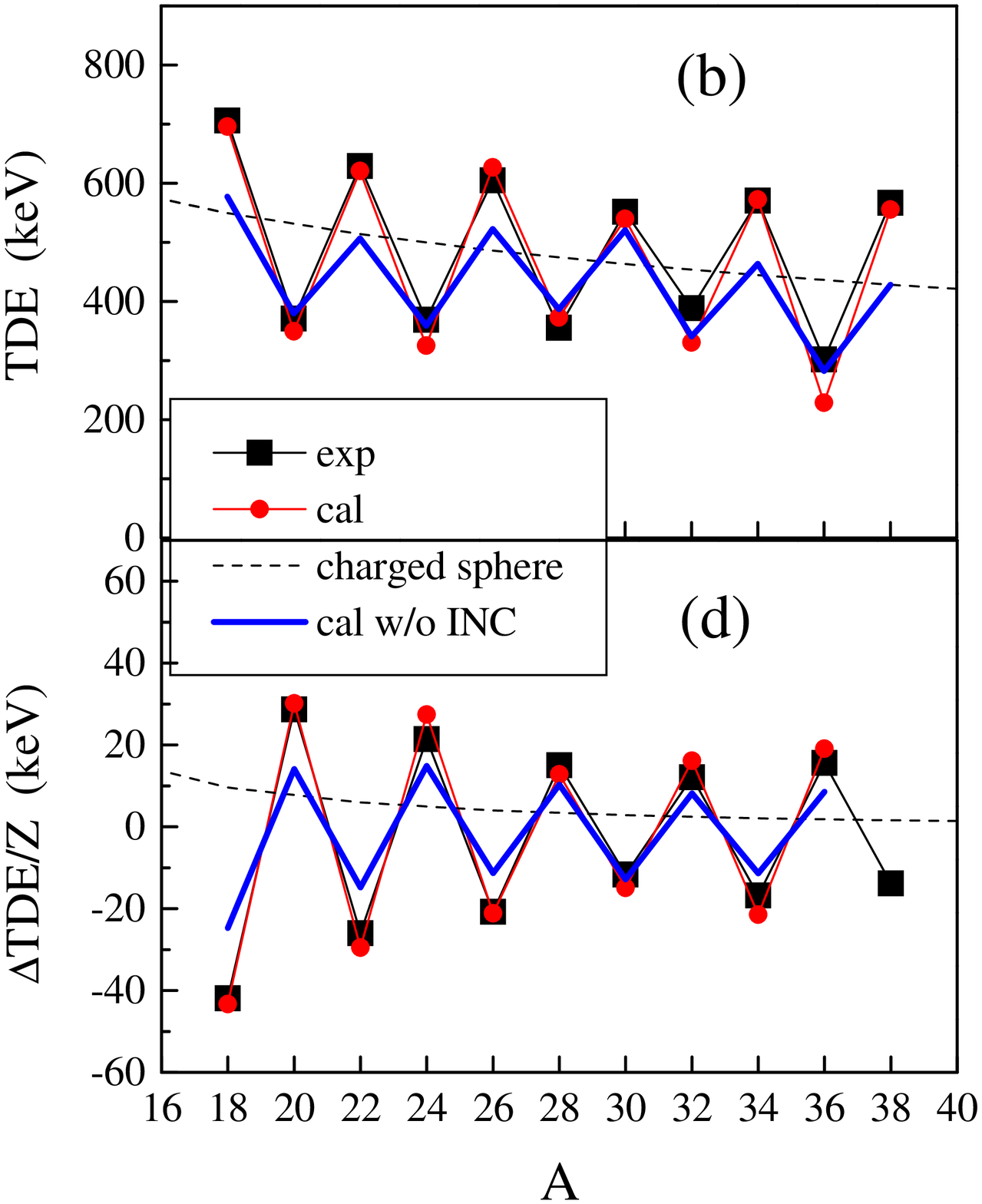}
  \caption{Experimental Coulomb displacement energy and triplet displacement energy.
(a) CDE's for $T=1/2$, (b) TDE's for $T=1$, (c) differences in CDE's
between $A$ and $A+2$ nuclei (shown as $\Delta$CDE/$Z$), and (d)
differences in TDE's between $A$ and $A+4$ nuclei (shown as
$\Delta$TDE/$Z$). Experimental data are taken from Ref.
\cite{Wang2012}. Theoretical curves from the Coulomb prediction
\cite{Bentley07} are shown for comparison.}
  \label{fig1}
\end{center}
\end{figure*}

In order to understand these anomalies in the superallowed Fermi
$\beta$-decay, we first investigate CDE's
\cite{Bentley07} and TDE's \cite{Kaneko13,Satula14} for ground and isobaric
analogue states in an isospin multiplet of the total isospin $T$, with
$T_{z}=(N-Z)/2$ specifying different states of isospin projection in the multiplet. CDE and TDE are defined as
\begin{eqnarray}  \label{eq:1}
 {\rm CDE}(A,T) & = & BE(T,T_{z>}) - BE(T,T_{z>}+p),  \nonumber\\
 {\rm TDE}(A,T) & = & BE(T,T_{z}=-1) + BE(T,T_{z}=+1) \nonumber \\
                & - & 2BE(T,T_{z}=0),
\end{eqnarray}
where $BE(T,T_{z})$ are the negative binding energies of ground and
isobaric analogue states. Here in the CDE's, $p$ protons exchange with neutrons, and
$T_{z>}$ is the $T_{z}$ for the larger-$Z$ isobar in a mirror pair. From the definitions in Eqs. (\ref{eq:1}), one sees that while the first quantity resembles differences in mirror binding energies,
the second is meaningful only for triplets. In our
notation, the total spin and parity $I^\pi$ of the isobaric
multiplets are implicitly included, but omitted in above equations
for simplicity.

In Fig. \ref{fig1}(a), the experimental
CDE's for the $sd$-shell with $T= -T_{z>}=+1/2, p=1$ are shown, where the binding energies
are those of the ground states of a mirror pair with $T_{z}=\pm 1/2$ \cite{Wang2012}. The experimental TDE's for the $sd$-shell with $T=1$ are shown in Fig. \ref{fig1}(b), where $BE(T,T_{z}=0)$ in Eq.
(\ref{eq:1}) is taken as the binding energy of the isobaric analogue state and
$BE(T,T_{z}=\pm 1)$ are those of the ground states. Both CDE's in Fig. \ref{fig1}(a)
and TDE's in Fig. \ref{fig1}(b) are compared with the Coulomb energy prediction with spherical
charged distribution (dotted curve) \cite{Bentley07}. One can see that the results of the Coulomb contribution for simple homogeneous spherical charge lie far above the CDE data in Fig. \ref{fig1}(a), and in Fig. \ref{fig1}(b-c), the results of charged sphere give only a smooth curve that tends to be the average of the staggerings seen from data.

In order to examine the characteristic patterns of CDE and TDE more
clearly, we use quantities measuring differences in CDE's between
$A$ and $A+2$ and in TDE's between $A$ and $A+4$ nuclei \cite{Kaneko13},
\begin{eqnarray}\label{eq:2}
 \Delta{\rm CDE}(A,T) = {\rm CDE}(A+2,T) - {\rm CDE}(A,T),  \nonumber\\
 \Delta{\rm TDE}(A,T) = {\rm TDE}(A+4,T) - {\rm TDE}(A,T).
\end{eqnarray}
In Fig. \ref{fig1}(c), one begins to see an odd-even staggering pattern in
$\Delta{\rm CDE/Z}$, while the Coulomb energy prediction gives only
a smooth average curve. Here by `odd' and `even'
we refer to the larger proton number ($Z>$ in Eq. (\ref{eq:1})) in a mirror pair with $A$. The occurrence of the odd-even staggerings can be explained by the differences between proton- and neutron-pairing
gaps \cite{Kaneko13}. In Fig. \ref{fig1}(b,d), data for TDE and $\Delta{\rm TDE/Z}$
exhibit strong zigzags, particularly strong for the lighter nuclei in the $sd$-shell.

In the following, we perform shell-model calculations with the
charge-independent interaction USDA \cite{Brown06} ($H_{0}$) and
additional isovector and isotensor $T=1,J=0$ INC interaction
($H_{INC}$). The INC force strengths are fitted to the experimental CDE's
and TDE's that are extracted from the binding-energy data of the $sd$-shell. After the experimental
CDE's and TDE's are described, we then
use the more sensitive quantities $\Delta{\rm CDE/Z}$ and $\Delta{\rm TDE/Z}$
to adjust the INC forces finely. The total Hamiltonian is given as
\begin{equation}  \label{eq:3}
 H = H_{0} + H_{INC},
\end{equation}
where $H_{INC}$ takes the form of a spherical tensor of rank two
\begin{equation}  \label{eq:4}
 H_{INC} = H'_{sp} + V_C + \sum_{k=1}^{2}V^{(k)}_{INC},
\end{equation}
with $V_C$ being the Coulomb interaction and $H'_{sp}$ the
single-particle Hamiltonian that includes the Coulomb
single-particle energies for protons and the single-particle energy
shifts $\varepsilon_{ls}$ due to the electromagnetic spin-orbit
interaction \cite{Andersson05}. The Coulomb single-particle energies
for protons are taken as $\varepsilon(d_{5/2})=3.60$,
$\varepsilon(s_{1/2})=3.55$, and $\varepsilon(d_{3/2})=3.60$ (all in
MeV) for the $sd$ model space. The two-body matrix elements of the
INC interaction $V^{(k)}_{INC}$ in Eq. (\ref{eq:4}), with $k=1$ and
$k=2$ for the isovector and isotensor component, respectively, are
related to those in the proton-neutron formalism
\cite{Ormand97,Bentley07} through
\begin{eqnarray}  \label{eq:5}
 V^{(1)}_{abcd,J} & = & V_{abcd,J}^{pp} - V_{abcd,J}^{nn}, \nonumber\\
 V^{(2)}_{abcd,J} & = & V_{abcd,J}^{pp} + V_{abcd,J}^{nn} - 2V_{abcd,J}^{pn},
\end{eqnarray}
where $V_{abcd,J}^{pp}$, $V_{abcd,J}^{nn}$, and $V_{abcd,J}^{pn}$
are, respectively, the $pp$, $nn$, and $pn$ matrix elements with
isospin $T=1$ and spin $J$. We note that more general two-body
charge-dependent interactions have been proposed in Refs. \cite{Ormand89,Nakamura94,Lam13}.

As our previous treatment for the upper $fp$ shell region
\cite{Kaneko12}, we adopt the INC isovector $V^{(1)}_{aabb,J=0}$ and
isotensor $V^{(2)}_{aabb,J=0}$ interactions in Eq. (\ref{eq:5}) with
$T=1,J=0$ for each orbit (here $a,b$ = $d_{5/2}$, $s_{1/2},
d_{3/2}$). Subsequently, only the diagonal interaction matrix
elements are considered and off-diagonal ones are neglected. We can
simply denote $V^{(1)}_{J=0}(a)=V^{(1)}_{abcd,J=0}$ and
$V^{(2)}_{J=0}(a)=V^{(2)}_{abcd,J=0}$ with $a=b=c=d$ in Eqs.
(\ref{eq:5}). The parameters are chosen to be (all in keV)
\begin{eqnarray}  \label{eq:6}
 V^{(1)}_{J=0}(d_{5/2})  =  100, \hspace{0.5 cm} V^{(2)}_{J=0}(d_{5/2})  =  150, \nonumber\\
 V^{(1)}_{J=0}(s_{1/2})  = -200, \hspace{0.5 cm} V^{(2)}_{J=0}(s_{1/2})  = -190, \\
 V^{(1)}_{J=0}(d_{3/2})  = -100, \hspace{0.5 cm} V^{(2)}_{J=0}(d_{3/2})  =
 170, \nonumber
\end{eqnarray}
so as to reproduce the experimental CDE and TDE data.

Calculations are performed with the shell model code MSHELL64
\cite{Mizusaki} for odd-mass nuclei with odd-neutron number and
isospin $T=$ 1/2, 3/2, 5/2, and 7/2, and for even-mass nuclei with
$T=$ 1, 2, and 3 in the $sd$ model space.

Figure \ref{fig1} shows the calculated results for $sd$-shell nuclei with mass
$A=18-39$. 
As one can see from Fig.
\ref{fig1}(a), the CDE curve calculated without the INC forces lies below the data points.
This reconfirms the well-known
"Nolen-Schiffer Anomaly" \cite{Nolen1969}, which states that there remains
a consistent under-estimate of the CDE's, even after the Coulomb
interaction and all the related corrections were taken into account. The calculations with the INC forces clearly shift the curve up to a point that the theoretical results agree nicely with data.
For Fig. \ref{fig1}(c), one may conclude that the $T=1,J=0$ isovector and
isotensor INC interactions are not very sensitive to the quantity $\Delta{\rm CDE}/Z$, because, as one can see from Fig. \ref{fig1}(c), the calculated patterns with and without the INC forces are not very different. However, TDE is found to be a more sensitive probe for the INC interactions. Figure \ref{fig1}(b) depicts that calculations without INC deviate significantly from the data, especially for the upper branch in the staggering pattern (for mass numbers 18, 22, 26, 34, and 38). By including
the INC interactions, an excellent agreement with the experimental
data for TDE's of the mass region $A=18-38$ can be obtained,
as seen in Fig. \ref{fig1}(b). In Fig.
\ref{fig1}(d), the experimental $\Delta{\rm TDE}/Z$ are also
better reproduced by considering the INC interactions.
The rms deviations between the calculated and experimental
CDE's and TDE's are 66 keV and 34 keV, respectively. Thus we have
demonstrated, consistent with previous findings
\cite{Zuker02,Kaneko12}, that the $T=1,J=0$ INC interactions are
important, and seem to be efficient, for
describing the characteristic behaviors in CDE's and TDE's of the $sd$-shell.
Nevertheless, we stress that INC forces with nonzero spin may be
needed for the other observations \cite{Bentley15}, which is the
central theme of the next discussion.

\begin{table*}[t]
\caption{Comparison between experimental and calculated CDE
         and TDE values (in keV) for the ground and excited multiplet states
         for $A=22,23,29-32$. All the CDE's
         are calculated for states with $T=1/2,T_{z>}=1/2,p=1$ (see Eq. (\ref{eq:1})), except for $I_{i}^{\pi}=3/2_{2}^{+}$ in $A=31$
         with $T=3/2,T_{z>}=1/2,p=1$.
         Experimental data are taken from Ref. \cite{Wang2012,ENSDF}.}
\begin{tabular}{cccc|cccc}   \hline \hline
        &  & CDE &  &  &  & TDE &    \\ \hline

\hspace{0.2cm} $A$ \hspace{0.2cm} & $I_i^\pi$ \hspace{0.3cm}
&  expt. \hspace{0.3cm}  & \hspace{0.3cm} calc. \hspace{0.3cm}
& \hspace{0.2cm} $A$ \hspace{0.2cm} & $I_i^\pi$ \hspace{0.3cm}
&  expt. \hspace{0.3cm}  & \hspace{0.3cm} calc. \hspace{0.0cm}   \\ \hline
23 & $3/2_1^+ $  & 4838.4 & 4798.1  & 22 & $0_1^+ $ & 620.1  &  620.6     \\
   & $5/2_1^+ $  & 4849.7 & 4806.5  &    & $2_1^+ $ & 555.7  &  530.0     \\
   & $7/2_1^+ $  & 4815.1 & 4806.6  &    & $4_1^+ $ & 461.1  &  424.6      \\
29 & $1/2_1^+ $  & 5724.8 & 5763.6  & 30 & $0_1^+ $ & 551.9  &  539.6     \\
   & $3/2_1^+ $  & 5835.0 & 5835.3  &    & $2_1^+ $ & 476.8  &  448.6     \\
   &             &        &         &    & $2_2^+ $ & 441.4  &  416.7     \\
31 & $1/2_1^+ $  & 6179.9 & 6196.0  & 32 & $1_1^+ $ & 388.2  &  332.0     \\
   & $3/2_1^+ $  & 6161.6 & 6179.1  &    & $2_1^+ $ & 328.4  &  310.6     \\
   & $5/2_1^+ $  & 6180.3 & 6168.6  &    & $0_1^+ $ & 298.4  &  256.6     \\
   & $3/2_2^+ $  & 6189.1 & 6137.4  &    &          &        &            \\  \hline \hline
\end{tabular}
\label{table1}
\end{table*}

To demonstrate that the present shell-model calculations with the
chosen effective INC interactions work also for the excited multiplet
states in this mass region, we list in Table \ref{table1} the CDE's and
TDE's for the ground and excited multiplet states in $A=22$, 23,
$29-32$. These mass numbers are chosen because they
involve the isotopes of our present interest ($^{31}$Cl, $^{32}$Cl,
and $^{23}$Al), for which transition strengths in
superallowed Fermi $\beta$-decay were observed \cite{Melconian11,Bennett16,Iacob06}, and these will be the main focus of our discussion in the later sections. For example,
the CDE of $A=31$ with $I_{i}^{\pi}=3/2_{2}^{+}$ shown in Table \ref{table1} is calculated for $T=3/2, T_{z>}=-1/2, p=1$
using Eq. (\ref{eq:1}), which involves the isobaric analogue state of the
daughter nucleus $^{31}$S in the Fermi $\beta$-decay of $^{31}$Cl, which will be discussed later in Fig. \ref{fig3}(b).
As one can see in Table \ref{table1}, the calculated CDE's and TDE's
are in a reasonably good agreement with the experimental data, except for $A=32$, where larger discrepancies are observed.

\begin{figure*}[t]
\includegraphics[totalheight=8.0cm]{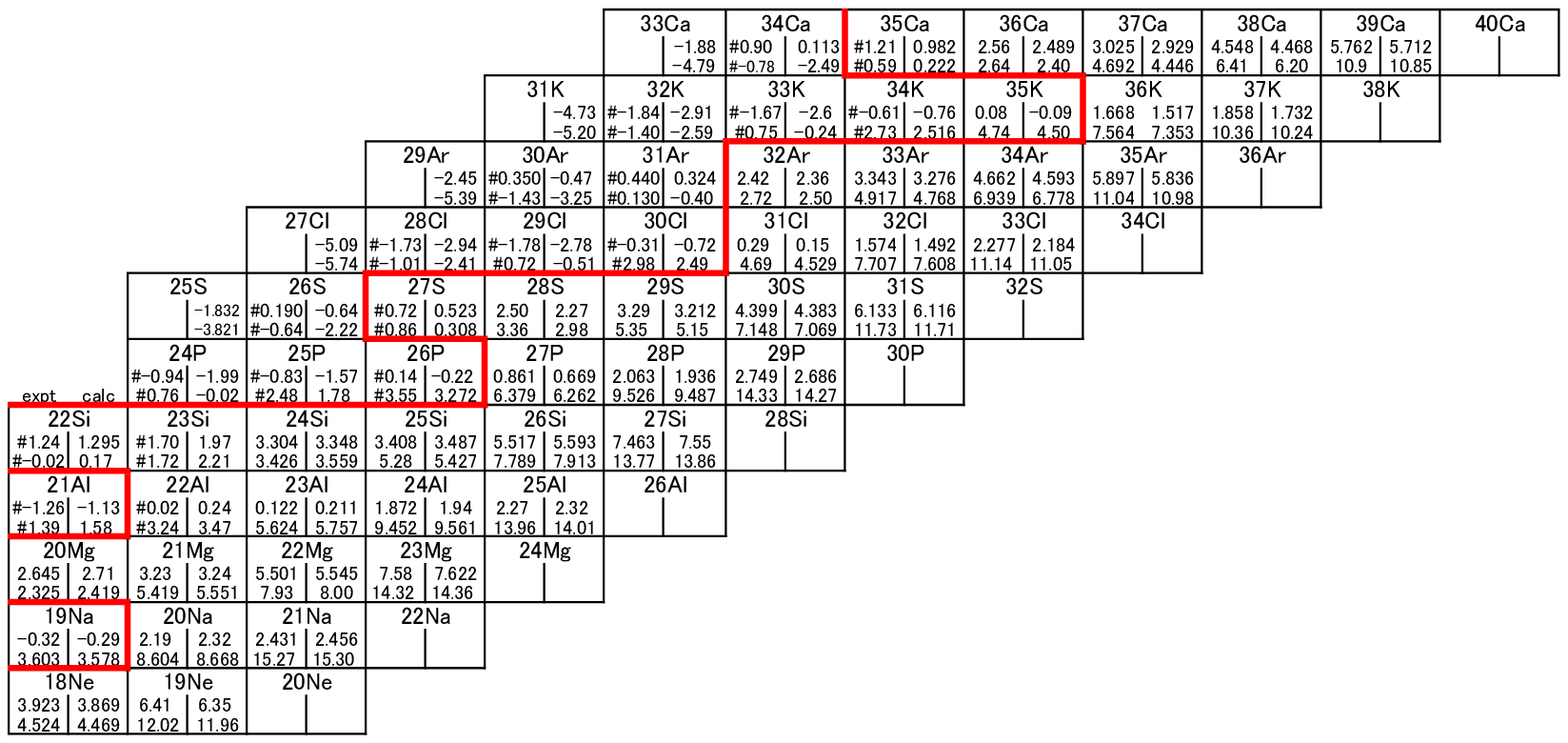}
\caption{(Color online) Calculated one- and two-proton separation
energies (in MeV) for odd-mass nuclei with isospin $T=1/2,3/2,5/2,7/2$ and
for even-mass nuclei with $T=1,2,3$. In each box, the first number
denotes one-proton separation energy and the second for two-proton
separation energy. For each isotope, the experimental data are
listed in the left box while the calculated results in the right.
The data are taken from Refs. \cite{Wang2012,ENSDF}, where the
``data" with \# are derived from systematic trends. Thick (red)
lines indicate the suggested proton drip-line.}
  \label{fig2}
\end{figure*}

By combining the shell-model CDE$(A,T)$ with the
experimentally-known $BE(A,T,T_{z}=T)$ of the neutron-richer isobars
\cite{Wang2012}, $BE(A,T,T_{z}=-T)$ of the corresponding proton-richer isobars can be calculated, and their one- and two-proton separation energies ($S_{p},
S_{2p}$) can thereby be obtained \cite{Kaneko13,Brown02}. Figure \ref{fig2}
shows a total number of 122 one- and two-proton separation energies
denoted in each box by the first and second numbers, respectively.
Among them, 72 have data to compare with and 50 are our predictions.
We have found that the calculated and experimental one- and
two-proton separation energies agree very well within a rms
deviation of about 94 keV and 148 keV, respectively. Without the
$T=1,J=0$ INC force in the calculation, the rms deviations for
$S_{p}$ and $S_{2p}$ increase by 18.5 keV and 30.9 keV,
respectively. It is to be noted that several proton-rich nuclei, for
example $^{26,27}$S, $^{30}$Ar, and $^{34,35}$Ca, however, exhibit
large discrepancies by about 200 keV. The thick (red) lines
represent the proton drip-line beyond which the one-proton and/or
two-proton separation energies become negative. Figure \ref{fig2}
also suggests several candidates for proton emitters in the
$sd$-shell mass region. Consistent with our results, two new
proton-unbound isotopes $^{30}$Ar and $^{29}$Cl have recently been
identified, pointing to a violation of isobaric symmetry in the
structure \cite{Mukha2015}.

Fermi $\beta$-decay provides one of the critical observations to
probe isospin symmetry, which leads to strict selection rules for
superallowed Fermi transition with the matrix element $M_{0}$
corresponding to the isospin symmetry. 
However, due to ISB, the Fermi matrix element for the
transition is modified as $|M_{F}|^{2}=|M_{0}|^{2}(1-\delta_{C})$,
where $\delta_{C}$, called isospin-breaking correction, is expected
to contain all the symmetry-breaking effects \cite{Towner08}, and is
known to be smaller than 2\%. For theoretical calculations, the
model space has to be very large because, in the first place, the
Coulomb force is long range in nature. Calculations in a
large shell-model space are not possible due to huge dimensions of
configuration. Hardy and Towner \cite{Hardy05,Towner08} thus wisely
divided $\delta_{C}$ into two parts,
$\delta_{C}=\delta_{C1}+\delta_{C2}$, where $\delta_{C1}$ arises
from configuration mixing in a restricted shell-model space and
$\delta_{C2}$ separately contributes from the mixing with outside,
which is estimated by computing radial overlap integrals with proton
and neutron radial functions \cite{Towner08}.

Recently, it has been reported \cite{Melconian11} that the
isospin-breaking correction $\delta_{C}$ = 5.3(9)\% for the observed
$ft=3200(30)$s in the Fermi transition from $^{32}$Cl to $^{32}$S is
anomalously larger than the typical values observed in many other
superallowed decays \cite{Towner08}. It has been suggested that the
closely-lying excited $T=1,I^{\pi}=1^{+}$ and $T=0,I^{\pi}=1^{+}$
states with the energy separation 188.2$\pm$1.2 keV
\cite{Melconian11} in $^{32}$S greatly enhances the isospin mixing,
and also the isospin-breaking correction \cite{Melconian11}. In
another very recent work \cite{Bennett16}, a large isospin mixing
has been observed in $^{31}$Cl $\beta$ delayed $\gamma$-decay
experiment. The investigation indicates a large isospin-mixing
between the $T=1/2$ and $T=3/2$ states with spin-parity
$I^{\pi}=3/2^{+}$. The matrix element of the Fermi transition
between $^{31}$Cl and $^{31}$S has been observed as $|M_{F}|^{2}$ =
2.4(1) for $T=3/2$ and $|M_{F}|^{2}$ = 0.48(3) for $T=1/2$, while
the symmetry-limit values are $|M_{0}|^{2}$ = 3 for $T=3/2$ and
$|M_{0}|^{2}$ = 0 for $T=1/2$, suggesting a considerable isospin
mixing. On the other hand, an early $^{23}$Al $\beta^{+}$-decay
experiment \cite{Iacob06} populated the $T=3/2$ IAS in $^{23}$Mg
with a log$ft$ of 3.31(3), corresponding to $|M_{F}|^{2}$ = 3.0. In
contrast to the two experiments \cite{Melconian11,Bennett16}
mentioned above, the $^{23}$Al $\beta^{+}$-decay result suggests no
isospin-mixing.

\begin{figure}[t]
\includegraphics[totalheight=7.0cm]{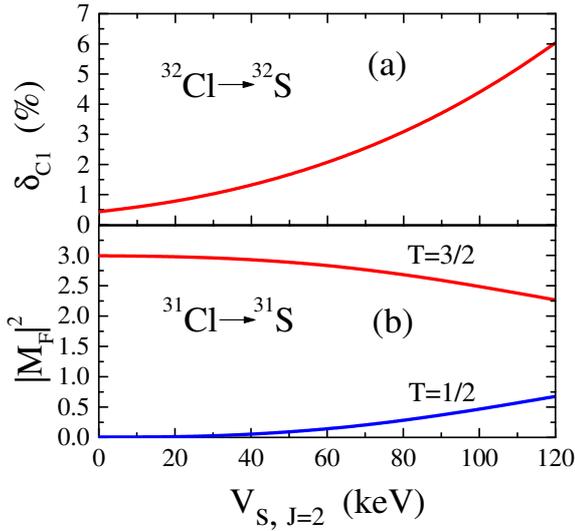}
  \caption{(Color online) Calculated isospin-breaking correction for
$^{32}$Cl$\rightarrow$ $^{32}$S and the Fermi transition
$|M_{F}|^{2}$ for $^{31}$Cl$\rightarrow$ $^{31}$S as a function of
the strength $V_{s, J=2}$.}
  \label{fig3}
\end{figure}

To understand these experimental data, Fermi matrix elements
$|M_{F}|^{2}$ are calculated. $\delta_{C1}$ can then be obtained
from $\delta_{C1}=1-|M_{F}|^{2}/|M_{0}|^{2}$, where $|M_{0}|^{2}=2$
and $|M_{0}|^{2}=3$ for $T=1$ and $T=3/2$, respectively. With the
$T=1,J=0$ INC interactions, the calculated Fermi transition matrix
elements for $^{32}$Cl $\rightarrow$ $^{32}$S and $^{31}$Cl
$\rightarrow$ $^{31}$S yield almost $|M_{F}|^{2}$ = 2 and 3,
respectively, very close to the symmetry-limit values. These results
are totally unexpected. Thus we must answer the question why the
superallowed Fermi transitions cannot be described by the $J=0$ INC
interactions for the $sd$ shell. In the $f_{7/2}$ shell, Bentley
{\it et al.} \cite{Bentley15} introduced a $J$-dependent INC
interaction in the investigation of mirror energy differences.
Moreover, there have been structure issues because
the calculations with the Woods-Saxon potential showed that only
two-body interactions related to the $s_{1/2}$ orbit should be
modified for the loosely-bound proton-rich nuclei
\cite{Ogawa99,Yuan14}. These previous investigations give us a hint
that $J$-dependent INC interactions related to the $s_{1/2}$ orbit
may be needed to explain the anomalous superallowed $\beta$-decay
data.

We now consider additional isovector and isotensor
INC interactions, with $T=1,J=1,2,3$, related to the $s_{1/2}$ orbit, and
assume a common strength with the same sign for simplicity (i.e.
$V^{(1)}_{sdsd,J}=V^{(2)}_{sdsd,J}= -2 V_{s,J}$). The $T=1$ INC interactions with
$J$ larger than 3 do not enter into the discussion simply because such
higher $J$ values are outside the couplings with the $s_{1/2}$ orbit
within the $sd$-shell. We found that only the
$J=2$ and 3 terms influence the superallowed
Fermi transition for $^{31}$Cl and $^{32}$Cl $\beta$-decays, while the contribution
from the $J=1$ term is negligibly small. Furthermore, our calculation shows
that to reproduce the experimental data of the Fermi transition
strengths $|M_{F}|^{2}$ for both $^{31}$Cl and $^{32}$Cl $\beta$-decays, the calculated results are insensitive to the combination of the $J=2$ and $J=3$ terms, and the parameters $V_{s,J=2}=100$ (in keV) and $V_{s,J=3}=0.0$ seem to reproduce the data. Therefore, we adopt only the $T=1,J=2$ INC
interaction for the present case, with a remark that further investigation with other $J$ values
may be necessary for other cases. Finally, we have
checked, and found that the effects of the additional isovector and
isotensor $T=1,J=2$ INC interactions on the CDE's and TDE's are
negligibly small. Therefore, the conclusions drawn earlier are not
affected.

Figure \ref{fig3} shows the isospin-breaking correction
$\delta_{C1}$ for the $^{32}$Cl $\beta$-decay and the Fermi
transition strengths $|M_{F}|^{2}$ for $^{31}$Cl $\beta$-decay as a
function of $V_{s}$. With zero $V_{s}$, $\delta_{C1}$ is small and
$|M_{F}|^{2}$ is 0.0 (3.0) for $T=1/2$ ($T=3/2$). The $\delta_{C1}$
value depends sensitively on $V_{s}$, and the $|M_{F}|^{2}$ for
$T=1/2$ ($T=3/2$) smoothly increases (decreases) with increasing
$V_{s}$. At $V_{s}$ = 100 keV, the isospin-breaking correction
becomes $\delta_{C1}$ = 4.39\%. By employing the value of
$\delta_{C2}$ = 0.865\% obtained with USDA by Melconian {\it et al.}
\cite{Melconian11}, the total isospin-breaking correction is
$\delta_{C}=\delta_{C1}+\delta_{C2}$ = 5.26\%. This value agrees
with the experimental data ($\delta_{C}$ = 5.3(9)\%) remarkably
well. For the $^{31}$Cl $\beta$-decay calculated with $V_{s}$ = 100
keV, our results show $|M_{F}|^{2}$ = 2.49 for $T=3/2$ and
$|M_{F}|^{2}$ = 0.47 for $T=1/2$, as seen in Fig. \ref{fig3} (b),
which are also in good agreement with the experiments. For the
$^{23}$Al $\beta^{+}$-decay, the calculated value is $|M_{F}|^{2}$ =
2.97, which indicates a log$ft$ of 3.32. Thus the same calculation
(with inclusion of the additional isovector and
isotensor $T=1,J=2$ INC interactions related to the $s_{1/2}$ orbit
and the same strength $V_s$) also reproduces correctly the
experimental suggested small isospin mixing in $^{23}$Al.

With the good description of the superallowed Fermi transition data,
we can further use the perturbation theory to estimate the isospin
admixture in a two-level mixing model, with the mixed states
expressed by
\begin{eqnarray}  \label{eq:7}
|\Psi_{a}\rangle = \hspace{0.3cm}{\rm cos}\theta |{\rm IAS} \rangle + {\rm sin}\theta |{\rm nonIAS} \rangle,  \nonumber \\
|\Psi_{b}\rangle = -{\rm sin}\theta |{\rm IAS} \rangle + {\rm
cos}\theta |{\rm nonIAS} \rangle.
\end{eqnarray}
In Eqs. (\ref{eq:7}), $|{\rm IAS} \rangle$ and $|{\rm nonIAS}
\rangle$ are the isobaric-analogue and non-isobaric-analogue states,
respectively, and $\theta$ is the mixing angle. Following Ref.
\cite{Tripathi13}, the mixing matrix element $v=\langle {\rm IAS}
|V_{\rm INC}| {\rm nonIAS} \rangle$ can be obtained as $v=\Delta E
{\rm sin}(2\theta)/2$ from ${\rm tan}^{2}\theta =
|M_{F}|_{b}^{2}$/$|M_{F}|_{a}^{2}$ and the energy separation $\Delta
E$ between IAS and non-IAS states. For the $^{31}$Cl decay, our calculated value $v=30$ keV
compares well with the experimental result $v=41$ keV. For the
$^{32}$Cl decay, our value $v=37.6$ keV agrees remarkably with the
experimental one $v=36.8$ keV. We can thus conclude that $T=1,J=2$
INC interactions with a single parameter can consistently be used to
understand the puzzling observations in superallowed Fermi
transition.

The above results have demonstrated the importance of nuclear
structure issues apart from the INC force itself. Nuclei with
different orbit-occupations may feel the INC force differently. In
the $^{31}$Cl and $^{32}$Cl $\beta$-decays, the occupation of
transformed protons in the $d_{3/2}$ orbit decreases due to the
$T=1,J=2$ INC interaction. This influences the Fermi beta-decay
transitions by suppressing the Fermi matrix elements $|M_F|^2$ from
3.0 to 2.49 for the $^{31}$Cl decay and from 2.0 to 1.91 for the
$^{32}$Cl decay. For the $^{23}$Al decay, however, the transformed
protons from the last occupied $d_{5/2}$ orbit are affected very
little by the $T=1,J=2$ INC interaction, and therefore, the ISB
effect becomes negligibly small. Thus the same calculation can
consistently explain why the ISB effect is large for $^{31}$Cl but
small for $^{23}$Al.

We point out that the present calculations involve free parameters.
Although the Coulomb single-particle energies and
each term in Eq. (\ref{eq:6}) have well-defined meanings, the
strengths are determined phenomenologically by fitting to data. It
is very much desired to understand the origin of the parameters,
which should be investigated with realistic forces that properly
contain the isospin-violating components.

In summary, we have investigated the effects of isospin
nonconserving forces on Coulomb displacement energy, triplet
displacement energy, and superallowed Fermi $\beta$-decay by
performing detailed shell-model calculations for the $sd$-shell
region. The isospin-invariant USDA effective interaction, together
with the Coulomb plus the $T=1,J=0$ isospin-nonconserving forces are
employed. We have shown that with a few fitted INC strengths, the
experimental CDE and TDE data of the entire mass region can be
described with high accuracy. We conclude that the INC force is
important for the $sd$-shell and inclusion of the $T=1,J=0$ INC
force seems to be sufficient for the description of CDE and TDE.
Based on the CDE calculation, we have further calculated a total of
122 one- and two-proton separation energies of the $sd$-shell mass
region. We have explicitly shown the location of proton drip-lines
and suggested potential candidates for proton emitters. However, the
calculation with the zero-spin INC force could not explain the three
superallowed Fermi $\beta$-decay experiments in $^{31}$Cl,
$^{32}$Cl, and $^{23}$Al \cite{Melconian11,Bennett16,Iacob06}. We
have demonstrated finally that these existing anomalies found in the
superallowed $\beta$-decay experiments can only be understood with
additional $T=1,J=2$ INC interaction that acts between the $d_{5/2}$
proton and the $s_{1/2}$ neutron orbit. We have found that the
calculated mixing matrix elements by shell model are in good
agreement with the experimentally extracted ones for the $^{31}$Cl
and $^{32}$Cl decays, and thus we conclude that the anomalous
behaviors in the superallowed $\beta$-decay can be understood by the
level mixing due to the $T=1,J=2$ INC force.

Research at Shanghai Jiao Tong University was supported by the
National Key Research and Development Program (No. 2016YFA0400501),
the National Natural Science Foundation of China (No. 11575112), and
the 973 Program of China (No. 2013CB834401).



\end{document}